\documentclass{svproc}
\usepackage{url}
\usepackage{xcolor}

\begin{document}
\mainmatter              
\title{A new approach to analysis of 2D  higher order quantum superintegrable systems}
\titlerunning{Quantum superintegrable systems }  
%
\author{Bjorn K. Berntson\inst{1} \and Ian Marquette\inst{2}
 \and Willard Miller, Jr.\inst{3} }
\authorrunning{Bjorn K. Berntson et al.} 
%
\tocauthor{Ivar Ekeland, Roger Temam, Jeffrey Dean, David Grove,
Craig Chambers, Kim B. Bruce, and Elisa Bertino}
\institute{Department of Mathematics, KTH Royal Institute of Technology,
Stockholm, Sweden\\
\email{bbernts@kth.se}
\and
School of Mathematics and Physics, The University of Queensland,\\ Brisbane, Australia\\
\email{i.marquette@uq.edu.au},\\ WWW home page:
\texttt{https://smp.uq.edu.au/profile/211/ian-marquette}
\and
 School of Mathematics, University of Minnesota,\\con
Minneapolis, Minnesota, U.S.A.\\
\email{mille003@math.umn.edu},\\ WWW home page:
\texttt{http://www-users.math.umn.edu/~mille003}}

\maketitle              

\begin{abstract} We  revise a method by Kalnins, Kress and Miller (2010) for constructing a canonical form for symmetry operators of 
arbitrary order for the Schr\"odinger eigenvalue equation $H\Psi \equiv (\Delta_2 +V)\Psi=E\Psi$ on any 
2D Riemannian manifold, real or complex, that admits a separation of variables in some orthogonal 
coordinate system. Most of this paper is devoted to describing the method. Details will be provided elsewhere.  As examples we revisit the Tremblay and Winternitz derivation of
 the Painlev\'e VI potential for a 3rd order superintegrable  flat space system that separates  in polar coordinates and, as new results,  we show that the Painlev\'e VI potential also appears for a 3rd order superintegrable system on the 2-sphere that separates in spherical coordinates, as well as a 3rd order superintegrable system on the 2-hyperboloid that separates in spherical coordinates and one that separates in horocyclic coordinates. The purpose of this project is to develop tools for analysis and classification of higher order superintegrable systems on any 2D Riemannian space, not just Euclidean space.
\end{abstract}
\section{Introduction}
In the paper \cite{KKM2010} the authors constructed a canonical form for symmetry operators of any order in 2D and used it to give the first  proof of the superintegrability  of the  quantum Tremblay, Turbiner,
 and Winternitz (TTW) system \cite{TTW} in polar coordinates, for all rational values of the parameter $k$. In the original method the various potentials were given and the problem was the construction of higher order symmetry operators that would verify superintegrability. The method was highly algebraic and required the solution of systems of difference equations on a lattice. Here, we consider   an arbitrary space admitting a separation in some orthogonal coordinate system (hence admitting a 2nd order symmetry operator), and search for all potentials $V$ for which the Schr\"odinger  equation admits an additional  independent symmetry operator of order higher than 2. Now the problem reduces to solving a system of partial differential equations.
 
 We give a brief introduction to the method and then specialize it to 3rd order superintegrable systems where we treat a few examples. We revisit the Tremblay and Winternitz derivation of
 the Painlev\'e VI potential for a 3rd order superintegrable  flat space system that separates  in polar coordinates, \cite{TW}, and we show among other new  results that the Painlev\'e VI potential also appears for a 3rd order superintegrable system on the 2-sphere that separates in spherical coordinates, as well as a 3rd order superintegrable system on the 2-hyperboloid that separates in spherical coordinates.

\section{The canonical form for a symmetry operator}
We consider a Schr\"odinger equation on a 2D real or complex Riemannian manifold with Laplace-Beltrami operator $\Delta_2$ and potential $V$:
\begin{equation} \label{TIS}H\Psi\equiv (-\frac{\hbar^2}{2}\Delta_2+V)\Psi=E\Psi \end{equation}
that also  admits an orthogonal  separation of variables. 
If $\{u_1,u_2\}$ is the  orthogonal separable coordinate system   the corresponding Schr\"odinger
operator can always be put in the  form 
\begin{equation} \label{canform}
H=-\frac{\hbar^2}{2}
\Delta_2+V(u_1,u_2)=
\frac{1}{f_1(u_1)+f_2(u_2)}\left(-\frac{\hbar^2}{2}\partial^2_{u_1}-\frac{\hbar^2}{2}\partial^2_{u_2}+v_1(u_1)+v_2(u_2)
\right)
\end{equation}
and, due to the separability, there is the second-order symmetry operator
\[
L_2= \frac{f_2(u_2)}{f_1(u_1)+f_2(u_2)}\left(-\frac{\hbar^2}{2}\partial^2_{u_1}+v_1(u_1)\right)-\frac{f_1(u_1)}{f_1(u_1)+f_2(u_2)}\left(-\frac{\hbar^2}{2}
\partial^2_{u_2}+v_2(u_2)\right),
\]
i.e., $
[H,L_2]=0$.
We  look for a  partial differential symmetry   operator of arbitrary order  ${\tilde L}(H,L_2,u_1,u_2)$ that satisfies 
\begin{equation}
[H,{\tilde L}]= 0.\label{newconditions1}
\end{equation}
We require that the symmetry operator take the standard form
\begin{equation}\label{standardLform}
{\tilde
L}=\sum_{j,k}\left(A^{j,k}(u_1,u_2)\partial_{u_1u_2}-B^{j,k}(u_1,u_2)\partial_{u_1}
-C^{j,k}(u_1,u_2)\partial_{u_2}+ D^{j,k}(u_1,u_2)
\right)H^jL_2^k.
\end{equation}
This can always be done.
Note that if the formal operator  $\tilde L$
contained partial
derivatives in $u_1$ and $u_2$ of orders $\ge 2$ we could 
 rearrange terms to achieve the
unique standard form (\ref{standardLform}).

Details of the derivation can be found in \cite{KKM2010}. 

Note that condition (\ref{standardLform}) makes sense, at least formally, for infinite order differential equations. Indeed, one can 
consider $H,L_2$ as parameters in these equations. Then once $\tilde L$ is
expanded as a power series in these parameters, the terms are reordered so that the powers of the parameters are on the right, before
they are replaced by explicit differential operators. Of course (\ref{standardLform}) is defined rigorously for finite order symmetry operators.

In this view we can write
\begin{equation}\label{generalLform}
{\tilde
L}(H,L_2,u_1,u_2)=A(u_1,u_2)\partial_{u_1 u_2}-B(u_1,u_2)\partial_{u_1}
-C(u_1,u_2)\partial_{u_2}
+ D(u_1,u_2),
\end{equation}
and consider $\tilde  L$ as an at most second-order  order differential operator in $u_1,u_2$ that is analytic in the parameters $H,L_2$. 
Then the above system of equations can be written in the more compact form {\small
\begin{equation} \label{partialxyHL}
\partial_{u_1}^2A+\partial_{u_2}^2A-2\partial_{u_2}B-2\partial_{u_1}C
=0,
\end{equation}
\begin{equation} \label{partialxHL}
\frac{\hbar^2}{2}(\partial_{u_1}^2B+\partial_{u_2}^2B)-2\partial_{u_2}A\,v_2-\hbar^2\partial_{u_1}D-Av'_2+(2\partial_{u_2}A\,f_2+Af'_2)H-2\partial_{u_2}A\,L_2=0,
\end{equation}
\begin{equation} \label{partialyHL}
\frac{\hbar^2}{2}(\partial_{u_1}^2C+\partial_{u_2}^2C)-2\partial_{u_1}Av_1-\hbar^2\partial_{u_2}D-Av'_1+(2\partial_{u_1}A\,f_1+Af'_1)H+2\partial_{u_1}A\,L_2=0,
\end{equation}
\begin{equation}\label{constanttermHL}
-\frac{\hbar^2}{2}(\partial_{u_1}^2D+\partial_{u_2}^2D)+2\partial_{u_1}B\,v_1+2\partial_{u_2}C\,v_2+Bv'_1+Cv'_2
\end{equation}
\[
-(2\partial_{u_1}B\,f_1+2\partial_{u_2}C\,f_2+Bf'_1
+Cf'_2)H+(-2\partial_{u_1}B+2\partial_{u_2}C)\,L_2=0.
\] }
We can view (\ref{partialxyHL}) as an equation for $A,B,C$ and (\ref{partialxHL}), (\ref{partialyHL}) as the defining equations for $\partial_{u_1}D, \partial_{u_2}D$.
Then $\tilde L$ is $\hat L$ with the terms in $H$ and $L_2$ interpreted as (\ref{standardLform}) and considered as partial differential operators.

We can simplify this system by noting that there are two functions\hfill\break $F(u_1,u_2,H,L_2)$, $G(u_1,u_2,H,L_2)$ such
that (\ref{partialxyHL}) is satisfied by 
\begin{equation}\label{ABCtoFG}
A=F,\qquad B=\frac12 \partial_{u_2}F+\partial_{u_1}G,\qquad C=\frac12\partial_{u_1} F-\partial_{u_2}G,
\end{equation}
Then the integrability condition for (\ref{partialxHL}), (\ref{partialyHL}) is (with the shorthand $\partial_{u_j}F=F_j$, $\partial_{u_j}\partial_{u_\ell}F=F_{j\ell}$, etc., for $F$ and $G$),{\small
\begin{eqnarray}- \hbar^2 G_{1222}-\frac14 \hbar^2 F_{2222}+2F_{22}(v_2-f_2H+L_2)
+3F_{2}(v'_2-f_2'H)+F(v''_2-f''_2H)&=&\nonumber\\  
 \hbar^2 G_{1112} -\frac14 \hbar^2 F_{1111}+2F_{11}(v_1-f_1H-L_2)
+3F_{1}(v'_1-f'_1H)+F(v''_1-f''_1H),&&\label{eqn1} \end{eqnarray} }

and equation (\ref{constanttermHL}) becomes {\small
\[ 
\frac14 \hbar^2 F_{1112}-2F_{12}(v_1-f_1H)-F_{1}(v'_2-f'_2H)+\frac14 \hbar^2
G_{1111}-
2G_{11}(v_1-f_1H-L_2)\] 
\begin{equation}\label{eqn2}-G_{1}(v'_1-f'_1H)=-\frac14\hbar^2F_{1222}+2F_{12}(v_2-f_2H)\end{equation}
$$+F_{2}(v'_1-f'_1H)+\frac14 \hbar^2
G_{2222}-
2G_{22}(v_2-f_2H+L_2)-G_{2}(v'_2-f'_2H).$$}
We remark that any solution of (\ref{eqn1}), (\ref{eqn2}) with $A,B,C$ not identically $0$ corresponds to a symmetry operator that does not commute with $L_2$, hence is algebraically  independent of the symmetries $H, L_2$.

\section{3rd order superintegrability}
To illustrate how equations (\ref{eqn1}) and (\ref{eqn2}) can be used to find potentials for superintegrable systems, we  provide detailed derivations of the determining equations for 3rd order superintegrability.  First we note that the most general 3rd order operator must be of the form (\ref{standardLform}) with  {\small
\[ A=A^0(x,y),\quad B=B^0(x,y)+B^H(x,y) H+B^L(x,y)L,\]\[C=C^0(x,y)+C^H(x,y) H+C^L(x,y)L,\
 D=D^0(x,y)+D^H(x,y) H+D^L(x,y)L,\] }
 or, in view of (\ref{ABCtoFG}),
\begin{equation}\label{eqn3} F(x,y)=F^0(x,y),\quad G(x,y)=G^0(x,y)+G^H(x,y)H +G^L(x,y)L.\end{equation}

Substituting (\ref{eqn3}) into (\ref{eqn1}), (\ref{eqn2}) and noting that the coefficients of independent powers of $H$ and $L$ in these expressions must vanish, we obtain 9 equations, (the first 3 from (\ref{eqn1}) and the next 6 from (\ref{eqn2})):

\begin{eqnarray*}&0=&-6v_1'F^0_1+6v_2'F^0_2-4v_1F^0_{11}+4v_2F^0_{22}-2\hbar^2G^0_{1112}-2\hbar^2G^0_{1222}\\
&&+2F^0v_2''-2F^0v_1'',\\
&0=& F^0_{11}+F^0_{22},\\
&0=&-\hbar^2G^H_{1112}-\hbar^2G^H_{1222}+3f_1'F^0_1-3f_2'F^0_2+2f_1F^0_{11}-2f_2F^0_{22}-F^0f_2''+F^0f_1'',\\
&0=&v_2'F^0_1+v_1'F^0_2+v_1'G^0_1-v_2'G^0_2+2F^0_{12}v_2+2F^0_{12}v_1+2v_1G^0_{11}-2v_2G^0_{22}-\\
&&\frac14\hbar^2G^0_{1111}+\frac14\hbar^2G^0_{2222},\\ 
&0=&v_1'G^L_1-v_2'G^L_2+2v_1G^L_{11}-2G^0_{11}-2v_2G^L_{22}-2G^0_{22},\\
&0=& G^L_{11}+G^L_{22},\\
&0=& -f_2'F^0_1-f_1'F^0_2+v_1'G^H_1-f_1'G^0_1-v_2'G^H_2+f_2'G^0_2-2F^0_{12}f_2-2F^0_{12}f_1+2v_1G^H_{11}\\
&&-2f_1G^0_{11}-2v_2G^H_{22}+2f_2G^0_{22}-\frac14\hbar^2G^H_{1111}+\frac14\hbar^2G^H_{2222},\\
&0=&-f_1'G^L_{1}+f_2'G^L_{2}+2f_2G^L_{22}-2f_1G^L_{11}-2G^H_{11}-2G^H_{22},\\
&0=&-f_1'G^H_1+f_2'G^H_2+2f_2G^H_{22}-2f_1G^H_{11}.\\
\end{eqnarray*}

\section{Some examples (mostly new)} We are particularly interested in potentials with nonlinear defining equations. First, we show that we get the result of Tremblay and Winternitz \cite{TW} that the  quantum system separating in polar coordinates in 2D Euclidean space admits potentials that are expressed in terms of the sixth Painlev\'e transcendent or in terms of the Weierstrass elliptic function. To do this we must put the system in the canonical form (\ref{canform}). The separable polar coordinates are $(x,y)=(r\cos(\theta),r\sin(\theta))$. For the canonical form we use the coordinates $\{u_1,u_2\}$, where $r=\exp(u_1),\ \theta =u_2$. Thus, $f_1(u_1)=\exp(2u_1)$ and $f_2(u_2)=0$. We know that these extreme potentials can appear only if the potential depends  on the angular variable alone, so we set $v_1(u_1)=0$.  Since we want only systems that satisfy nonlinear equations alone, whenever an explicit linear equation for the potential appears, we require that it vanish identically. 

We obtain a solution 
\begin{eqnarray*} F^0 &=& 4\hbar^2\exp(-u_1)\sin(u_2),\quad G^L= -8 \exp(-u_1)\cos(u_2)+a_4 u_2+a_3,\\  G^0 &=& -U_1(u_2)\exp(-u_1)+U_2(u_2),\quad
 G^H=a_5,\end{eqnarray*}
subject to the conditions
\begin{eqnarray}\label{cond1}&0=&a_4\frac{dv_2}{du_2}+2\frac{d^2U_2}{du_2^2},\\ \label{cond2}&0=& \hbar^2\frac{d^4U_2}{du_2^4}+4a_4\frac{dv_2}{du_2}v_2-4\frac{dv_2}{du_2}\frac{dU_2}{du_2},\\
\label{cond3}&0=&8v_2\cos(u_2)+4\frac{dv_2}{du_2}\sin(u_2)-\frac{d^2U_1}{du_2^2}- U_1,\\
\label{cond4}&0=&\frac{dv_2}{du_2}\frac{dU_1}{du_2}-\hbar^2\frac{d^3v_2}{du_2^3}\sin(u_2)-4\hbar^2 \frac{d^2v_2}{du_2^2}\cos(u_2)\\
&&+2\sin(u_2)(\hbar^2+4v_2)\frac{dv_2}{du_2}-2v_2\left(2\hbar^2\cos(u_2)-8v_2\cos(u_2)+U_1\right) \nonumber
.\end{eqnarray}

There are basically two cases to consider: \begin{enumerate} \item $a_4=0$. \\ \medskip

Then condition (\ref{cond1}) says that $U_2$ is linear in $u_2$. Thus condition (\ref{cond2}) is a linear equation for $v_2(u_2)$ which must vanish. Then condition (\ref{cond3}) can be solved for $U_1(u_2)$ and the result substituted into condition (\ref{cond4}) to obtain an equation for $v_2(u_2)$. After some manipulation (using the fact that $V_2$ is unchanged under transformations $W\to W+c$, where $c$ is a constant), we obtain an equation characterizing Painlev\'e VI, in agreement with \cite{TW}, equation (4.27):
\begin{equation} \label{PVI} \hbar^2\left(\sin(u_2) \frac{d^4W}{du_2^4}+4\cos(u_2)\frac{d^3W}{du_2^3}-6\sin(u_2)\frac{d^2W}{du_2^2}-4\cos(u_2)\frac{dW}{du_2}\right)   \end{equation}
\[ -12\sin(u_2)\frac{dW}{du_2}\frac{d^2W}{du_2^2}-4\cos(u_2)W\frac{d^2W}{du_2^2}-4(\beta_1\sin(u_2)-\beta_2\cos(u_2))\frac{d^2W}{du_2^2}\]
\[-16\cos(u_2)\bigg(\frac{dW}{du_2}\bigg)^2+8\sin(u_2)W\frac{dW}{du_2}-8(\beta_1\cos(u_2)+\beta_2\sin(u_2))\frac{dW}{du_2}=0.\] Here $v_2(u_2)=\frac{dW(u_2)}{du_2}$.
\item $a_4 \ne 0$. \medskip\\   Solving condition (\ref{cond1}) for $v_2(u_2)$ and substituting the result and (\ref{cond1}) into (\ref{cond2}) we obtain the equation that characterizes the Weierstrass $\wp$-function (in fact it is a translated and rescaled version):
\begin{equation} \label{Weierstrass} \hbar^2 \frac{d^3v_2}{du_2^3}-12\frac{dv_2}{du_2}v_2+12a_1\frac{dv_2}{du_2}=0.\end{equation} Thus $v_2(u_2)=\hbar^2\wp( u_2-u_{2,0};g_2,g_3)+a_1$, where $u_{2,0}$, $g_2$, and $g_3$ are arbitrary constants. As shown in \cite{TW} this solution is subject to the compatibility conditions (\ref{cond3}) and (\ref{cond4}), which leads to a complicated nonlinear differential equation for $v_2(u_2)$.
\end{enumerate}

With this verification out of the way, we consider the analogous system on the 2-sphere, separable in spherical coordinates. Here $s_1=\sin(\theta)\cos(\phi),\ s_2=\sin(\theta)\sin(\phi),s_3=\cos(\theta)$ with $s_1^2+s_2^2+s_3^2=1$. This system is in canonical form with coordinates $u_1,u_2$ where
\begin{equation}\label{cansp} \sin(\theta)=(\cosh(u_1))^{-1},\ \phi=u_2,\ f_1(u_1)=(\cosh(u_1))^{-2},\ f_2(u_2)=0. \end{equation} As before we look for solutions such that $v_1(u_1)=0$ and $v_2$ satisfies a nonlinear equation only.

The computation is very similar to that for the Euclidean space example. We obtain the solution 
\[ F^0 = 4\hbar^2\cosh(u_1)\sin(u_2),\ G^L= 8\sinh(u_1)\cos(u_2)+a_{4}u_2+a_3,\]
\[G^0 =  \sinh(u_1)\ U_1(u_2)+U_2(u_2),\
 G^H=a_5,\]
 subject to the conditions (\ref{cond1})-(\ref{cond4}), exactly the same as for Euclidean space. 
  Thus the system on the 2-sphere also admits Painlev\'e VI and special Weierstrass potentials for 3rd order superintegrability. It is clear from these results that these systems in Euclidean space can be obtained as B\^ocher contractions, \cite{Bocher}, chapter 15, of the corresponding systems on the 2-sphere. 

\bigskip

Next we consider spherical coordinates  on the hyperboloid $s_1^2-s_2^2-s_3^2=1$,
\[ s_1=\cosh(x),\ s_2=\sinh(x)\cos(\phi),\ s_3=\sinh(x)\sin(\phi).\]
For the canonical form we find  
\[ \tanh \bigg(\frac{u_1}{2}\bigg)=\exp(x),\ u_2=\phi,\quad  f_1(u_1)=(\sinh(u_1))^{-2},\ f_2(u_2)=0,\]
 and we look for solutions such that $v_1(u_1)=0$ and $v_2(u_2)$ satisfies only a nonlinear equation. We obtain the solution 
 \[ F^0 = 4\hbar^2\sinh(u_1)\sin(u_2),\ G^L= 8\cosh(u_1)\cos(u_2)+a_{4}u_2+a_3,\]
\[G^0 =  \cosh(u_1)\ U_1(u_2)+U_2(u_2),\
 G^H=a_5,\]
 subject to the conditions (\ref{cond1})-(\ref{cond4}), again exactly the same as for flat space. Thus the system on the 2-hyperboloid  admits Painlev\'e VI and special Weierstrass potentials for 3rd order superintegrability.
 
For our next example we consider horocyclic coordinates $\{u_1,u_2\}$ on the hyperboloid $s_1^2-s_2^2-s_3^2=1$, e.g. \cite{KKM2018}, section 7.7:
 \begin{equation}\label{horcoords} s_1=\frac12\bigg(u_1+\frac{u_2^2+1}{u_1}\bigg), \ s_2=\frac12\bigg(u_1+\frac{u_2^2-1}{u_1}\bigg), \ s_3=\frac{u_2}{u_1}.\end{equation}
 These coordinates are separable and the canonical system is defined by $f_1(u_1)=1/u_1^2$, $f_2(u_2)=0$. We look for systems such that $v_1(u_1)=0$, in analogy with our first three examples.

We obtain the solution 
\begin{eqnarray*}F^0 &=&-\frac12 a_8 \hbar^2u_1,\quad G^L= \frac{u_1^2(a_8 u_2+a_9)}{2}-\frac{a_8u_2^3}{6}-\frac{a_9u_2^2}{2}+a_{10}u_2,\\
G^0& =& \frac{u_1^2}{2}U_1(u_2)+U_2(u_2),\quad
 G^H=a_7,\end{eqnarray*}
 subject to the conditions 
  \begin{eqnarray}&0=& a_8v_2+2\frac{d2U_1}{du_2},\label{hcond1}\\
 &0=& \frac12\hbar^2 a_8\frac{d^3v_2}{du_2^3}-4a_8\frac{dv_2}{du_2}v_2+4\frac{dv_2}{du_2}\frac{dU_1}{du_2},\label{hcond2}\\
  &0=&(2a_{10}-2a_9 u_2 -a_8u_2^2)\frac{dv_2}{du_2}-4(a_9+a_8 u_2)v_2+4U_1+4\frac{d^2U_2}{du_2^2},\label{hcond3}\\
 &0=&   -2\hbar^2 a_8 u_2^2\frac{dv_2}{du_2}+16(a_9+a_8 u_2)v_2^2-4(2a_{10}-2a_9u_2+a_8 u_2^2)\frac{dv_2}{du_2}v_2 \label{hcond4} \\
   &&\quad +\frac{\hbar^2}{2}(2a_{10}-2a_9 u_2-a_8u_2^2)\frac{d^3v_2}{du_2^3}-4\hbar^2(a_9+a_8 u_2)   \nonumber \\
   &&\quad -16V_2U_1+8\frac{dv_2}{du_2}\frac{dU_2}{du_2}.   \nonumber \end{eqnarray}

There are again two basic cases here: \begin{enumerate} \item $a_8=0$. \\ 

Then conditions (\ref{hcond1}) and (\ref{hcond2}) say that $U_1$ is a constant: $U_1(u_2)=d_1$.  Then condition (\ref{hcond3}) can be solved for $U_2(u_2)$ and the result substituted into condition (\ref{hcond4}) to obtain an equation for $v_2(u_2)$:
\begin{equation}\label{horocyclicnonlinear}   -4a_9\bigg(\frac{dW}{du_2}\bigg)^2+\left((-3a_9u_2+3a_{10})\frac{d^2W}{du_2^2} +4d_1\right)\frac{dW}{du_2}+(-a_9 W+2d_1u_2-2d_3)\frac{d^2W}{du_2^2}\end{equation}
\[+\hbar^2a_9\frac{d^3W}{du_2^3}-\frac14\hbar^2(-a_9u_2+a_{10})\frac{d^4W}{du_2^4}=0,\qquad {\rm where}\quad v_2(u_2)=\frac{dW(u_2)}{du_2}.  \]

\item $a_8\ne 0$. \\ \medskip  Here we can solve (\ref{hcond1}) for $v_2(u_2)$ and 
substitute the result into (\ref{hcond2}) to obtain the equation
\begin{equation}\label{hcond5}\hbar^2\frac{d^3v_2}{du_2^3}-12v_2\frac{dv_2}{du_2}+12a_1\frac{dv_2}{du_2}=0. \end{equation}

and it follows that $v_2(u_2)=\hbar^2\wp( u_2-u_{2,0};g_2,g_3)+a_1,$
where $u_{2,0}$, $g_2$, and $g_3$ are arbitrary constants. 
\end{enumerate}

\section*{Acknowledgments}
We thank Pavel Winternitz for helpful discussions and Adrian Escobar for pointing out the 
relevance of the paper \cite{KKM2010} to classification of 3rd order superintegrable systems.
W.M. was partially supported by a grant from the Simons Foundation (\# 412351 to Willard Miller, Jr.). \ I.M. was supported by the Australian Research Council Discovery Grant DP160101376 and Future Fellowship FT180100099.

%
%

\end{document}